\documentclass[a4paper,english,conference]{IEEEtran}
\usepackage[T1]{fontenc}
\usepackage{color}
\usepackage{babel}
\usepackage{units}
\usepackage{amsmath}
\usepackage{graphicx}
\usepackage[unicode=true,
 bookmarks=true,bookmarksnumbered=true,bookmarksopen=true,bookmarksopenlevel=1,
 breaklinks=false,pdfborder={0 0 0},pdfborderstyle={},backref=false,colorlinks=false]
 {hyperref}
\hypersetup{pdftitle={Your Title},
 pdfauthor={Your Name},
 pdfpagelayout=OneColumn, pdfnewwindow=true, pdfstartview=XYZ, plainpages=false}

\makeatletter

\usepackage[caption=false,font=footnotesize]{subfig}
\usepackage{cite}

\usepackage{flushend}

\@ifundefined{showcaptionsetup}{}{%
 \PassOptionsToPackage{caption=false}{subfig}}
\usepackage{subfig}
\makeatother

\begin{document}
\title{On the Construction of Polar Codes in the Middleton Class-A Channels }
\author{\IEEEauthorblockN{Ammar Hadi\textcolor{black}{$^{*}$}, Emad Alsusa\textcolor{black}{$^{\ddagger}$},
and Khaled M. Rabie$^{\diamond}$\\
\textcolor{black}{$^{*}$}Wireless Communication Department, Ministry
of Communication, Iraq\\
\textcolor{black}{$^{\ddagger}$}School of Electrical and Electronic
Engineering, the University of Manchester, UK \\
$^{\diamond}$\textcolor{black}{Department of Engineering, Manchester
Metropolitan University, UK }\\
Email: ammar.hadi@scis.gov.iq, e.alsusa@manchester.ac.uk, k.rabie@mmu.ac.uk}}
\maketitle
\begin{abstract}
Although power line communication (PLC) systems are available everywhere,
unfortunately these systems are not suitable for information transmission
due to the effects of the impulsive noise. Therefore, many previous
studies on channel codes have been carried out for the purpose of
reducing the impulsive noise in such channels. This paper investigates
some methods for the construction of polar codes under PLC systems
in the presence of Middleton class-A noise. We discuss here the most
feasible construction methods which already have been adopted with
other channels. In addition, we present an illustrative example for
the construction in these methods and also we discuss a comparison
between the methods in terms of performance. \\
\end{abstract}


\section{Introduction}

PLC systems have drawn a huge interest as they can be easily accessed
over the existing power line networks, which are available in almost
every building on the planet. Hence, PLC technology is preferred in
some wireless environments which have a significant propagation loss.
Moreover, the huge demand for communication and internet networks
makes PLC one of the most important competing technologies. However,
PLC networks are not suitable for the communication signals because
the connected electrical appliances in these networks can be considered
as abundant sources of interference and noise \cite{Modeling&AnalysisofNoiseinPLC}
. Thus, the noise over the PLC channels could be divided into two
types, namely, colored background noise and impulsive noise. In fact
the latter has the worst impact on the PLC network reliability due
to its large power spectral density (PSD) which is normally $10-15$
dB higher than the PSD of the colored background noise. For the PLC
noise, Middleton class-A model is adopted in this paper because it
is the most accepted model amid PLC channel models \cite{MiddletonclassA}. 

Many studies have been proposed to mitigate the impact of impulsive
noise; this includes the use of nonlinear preprocessing and channel
coding \cite{Khaled1ReferencePaper2020,TurboPLC,TurbodecodinginIN,Khaled2ReferencePaper2020,2015PCoverPLCSimilarToOurWork}.
As for channel coding, binary turbo codes were proposed over the PLC
channels in order to reduce the impact of impulsive noise\cite{TurboPLC}
\cite{TurbodecodinginIN}. In this regard, the authors in \cite{2015PCoverPLCSimilarToOurWork}
have discussed the performance of polar codes over the PLC channels
by using signal limiter. The popular code family, which is the low
density parity check (LDPC), was investigated with both single-carrier
and multi-carriers techniques over the PLC channels and resulted in
further enhancement in the reduction of impulsive noise impact\cite{DecodingLDPCoverINNakagawa,LDPCwithPLCConference,ImprovingHomePlugPLCwithLDPC}.
Furthermore, comparisons between the performance of LDPC and turbo
codes in the context of the PLC channels were explored in \cite{2014ComparisionLDPC&TCoverPLC}.
In the other hand, previous research has presented a comparison between
polar codes with LDPC codes in both single-carrier and multi-carriers
modulations over PLC channels \cite{MyPaperPraguePLC}. Also, a recent
work showed a comparison between these two codes for smart grid (SG)
using orthogonal frequency modulation multiplexing (OFDM) in the presence
of impulsive noise \cite{PolarCodeWithPLC2019}. 

In the last decade, polar codes family has attracted an attention
since it can achieve the maximum capacity of the channel by using
a low complexity decoder in comparison to some other family codes
\cite{AmmarHadiMHD}. The underlying key of polar codes is the channel
polarization phenomenon in which a number of the binary discrete memoryless
channels (B-DMCs) are combined to create one comprehensive channel.
Then, this single channel splits into two types of bit-channels namely,
information, and frozen bit-channels. The frozen bits are fixed to
zero and the decoder has the full knowledge about their positions
in the codeword \cite{AmmarHadiOSDD}. 

Unlike other channel codes, polar codes are not universal codes which
mean that the code construction is dependent on the signal to noise
ratio (SNR) of the channel. It is worth mentioning that the heuristic
method was applied as the earliest method of construction and it was
utilized by the help of the binary erasure channels (BEC) \cite{ArikanletterRMandPC}.
The key element of the heuristic method is the Bhattacharyya parameter
which can be denoted by $Z(W)$. The definition of $Z(W)$ is the
upper bound of the error rate for a channel in binary transmission.
The initial value $Z(W_{1})$ for of this method is usually considered
as $0.5$. On the other hand, the heuristic based capacity method
was adopted in polar codes in \cite{APracticalConstructionPCinAWGN}.
This method differs from the normal heuristic method in its initial
value where $Z(W_{1})$ is evaluated according to the capacity of
the using channel. Moreover, Monte Carlo method which based on repeated
random sampling for achieving results \cite{PolarCode}. It should
be noted that there are other construction methods have been developed
such as quantization approximations method \cite{HowtoconstructPC},
and a method based on linear complexity convolution \cite{MoriTanakaPerformanceofPC}.
Furthermore, a practical construction method was proposed for the
general binary input channels by introducing a new algorithm for finding
the exact values of the Bhattacharyya parameter bounds \cite{APracticalConstructionMethodforPC}.
Also, a comparison between the construction methods is presented for
the purpose of finding the best design-SNR for the different methods
in additive white Gaussian noise (AWGN) channel \cite{AcomparitveStudyofPCConsofAWGNArXiv}. 

In this paper, we investigate methods for the construction of polar
codes over PLC channels under Middleton class-A model. Therefore,
we focus on three methods which are heuristic, heuristic based capacity,
and Monte Carlo as they are feasible in the PLC channels. In addition,
we present an illustrative example to show how the change in SNR affects
on the construction according to each method. For a practical comparison,
we show the performance in terms of error rate ratio versus SNR. 

The rest of this paper is organized as follows. In Section II, briefly
review the preliminaries. In Section III, the Bhattacharyya parameter
was discussed in the Middleton class-A channel. The three methods
of the construction are discussed in Section IV. The simulation results
with discussion are given in Section V. Finally, conclusions are drawn
in Section VI.

\section{Preliminaries}

\subsection{Middleton's Class-A Noise Model}

Middleton class-A has been extensively used as a proper model for
the PLC systems due to its suitability for the impulsive noise characterization
over PLC. The model contains background noise and impulsive noise.
In general, the probability density function (PDF) for this model
is expressed as\textcolor{black}{{} \cite{MiddletonclassA}}

\begin{equation}
P_{Z}(z)={\displaystyle \sum_{m=0}^{\infty}\exp(-A)\frac{A^{m}}{m!}}\frac{1}{\sqrt{2\pi}\sigma_{m}}\exp\left(-\frac{z^{2}}{2\sigma_{m}^{2}}\right),\label{eq:PDFImulsive}
\end{equation}
where

\begin{equation}
\sigma_{m}=\sqrt{\sigma^{2}\frac{\frac{m}{A}+\Gamma}{1+\Gamma}},
\end{equation}
where $A$ is the impulsive index, $\Gamma$ is the Gaussian-to-impulsive
power ratio given as $\Gamma=\nicefrac{\sigma_{G}^{2}}{\sigma_{I}^{2}}$,
$\sigma_{G}^{2}$ is the Gaussian noise variance, $\sigma_{I}^{2}$
is the impulsive noise variance, and the variance of the total noise
is given by $\sigma=\sqrt{\sigma_{G}^{2}+\sigma_{I}^{2}}$. Note that
the impulsive index identifies the average number of impulses over
the signal period, and $\Gamma$ indicates the strength of impulsive
noise compared to the background noise. It was found in \cite{TheoreticalPLCequation}
that the probability of errors for Middleton's class-A channel in
the binary phase shift keying (BPSK) modulation scheme is 

\begin{equation}
P_{e}=\frac{\exp(-A)}{2}\sum_{m=0}^{\infty}\frac{A^{m}}{m!}erfc\Biggl(\sqrt{\frac{(1+\Gamma)}{\left(\frac{m}{A}+\Gamma\right)}\frac{E_{b}}{N_{0}}}\Biggr),\label{eq:TheoryPLC}
\end{equation}
where $erfc(.)$ denotes the complementary error function. 

It should be pointed out that the capacity of the Middleton class-A
is given as \cite{ChannelCapacityMiddletonOrig} 

\begin{equation}
C=\sum_{m=0}^{M}\pi_{m}C_{m},
\end{equation}
where $C_{m}$ is the capacity of the AWGN channel capacity with state
$m$ and it can be given by 

\begin{equation}
C_{m}=B\log_{2}(1+\frac{s}{n}),
\end{equation}
where $B$ is the bandwidth, $s$ is the total signal power, and $n$
is the power of the noise. It is worth mentioning that the capacity
of the Middleton class-A channel approaches the AWGN capacity with
low impulsive noise.

\subsection{Polar Codes }

In polar codes, $N$ number, usually power of $2$, of identical copies
of a channel $W$ are combined to become one channel $W_{N}:X_{N}\rightarrow Y_{N}$,
where $X_{N}=(x{}_{0},x_{1},...,x_{N-1})$ and $Y_{N}=(y_{0},y_{1},...,y_{N-1})$
are the corresponding channel inputs and channel outputs vectors,
respectively. The channel $W_{N}$ can be divided into two different
sets by the polarization phenomenon. The information set includes
the noiseless bit-channels which are useful for transmission, while
the bit-channels in the frozen set are fixed to zero values by the
encoder of polar codes which utilized by

\begin{equation}
X=UG_{N},\label{eq:1-1}
\end{equation}
where $G_{N}$ denotes the generator matrix. It should be mentioned
that $U$ vector includes the entirely information and frozen bits. 

\section{Bhattacharyya parameter of Middleton class-A}

As we mentioned above that the Bhattacharyya parameter is an important
tool for measuring the reliability of channels and it can be given
as 

\begin{equation}
Z(W)=\sum_{y\in Y}\sqrt{p(y|0)p(y|1)}.\label{eq:BhataEq}
\end{equation}

For the binary erasure channel (BEC), applying (\ref{eq:BhataEq})
resulted in $Z(w)=\epsilon$, where $\epsilon$ is the erasure probability.
In a similar way, the Bhattacharyya parameter for the binary symmetric
channel (BSC) is equivalent to $\sqrt{p(1-p)}$, where $p$ is the
crossover probability. 

Furthermore, $Z(W)$ can be given for any continuous channel by 

\begin{equation}
Z(W)=\int_{-\infty}^{\infty}\sqrt{p(y|0)p(y|1)}dy.\label{eq:BhatCont}
\end{equation}
Therefore, in the practical AWGN channel, it can be calculated by
\cite{APracticalConstructionPCinAWGN} 
\begin{eqnarray}
Z(W) & = & \int_{-\infty}^{\infty}\sqrt{\frac{1}{2\pi\sigma^{2}}e^{\frac{-(y-1)^{2}}{2\sigma^{2}}}\frac{1}{2\pi\sigma^{2}}e^{\frac{-(y+1)^{2}}{2\sigma^{2}}}}dy\label{eq:Z(W)forAWGN}\\
 & = & e^{-\frac{1}{2\sigma^{2}}},
\end{eqnarray}
\begin{figure*}[tbh]
\begin{align}
Z(W) & =\int_{-\infty}^{\infty}\sqrt{\sum_{m=0}^{\infty}\frac{A^{m}}{m!}\frac{1}{\sqrt{2\pi}\sigma_{m}}\exp(-\frac{(y+1)^{2}}{2\sigma_{m}^{2}})\sum_{m=0}^{\infty}\frac{A^{m}}{m!}\frac{1}{\sqrt{2\pi}\sigma_{m}}\exp(-\frac{(y-1)^{2}}{2\sigma_{m}^{2}})}dy.\label{eq:BhataIN}
\end{align}

\rule[0.5ex]{1\linewidth}{0.5pt}
\end{figure*}
\begin{figure*}[tbh]
\subfloat[Erasure rate $=0.4$ ]{\begin{centering}
\includegraphics[width=0.65\columnwidth]{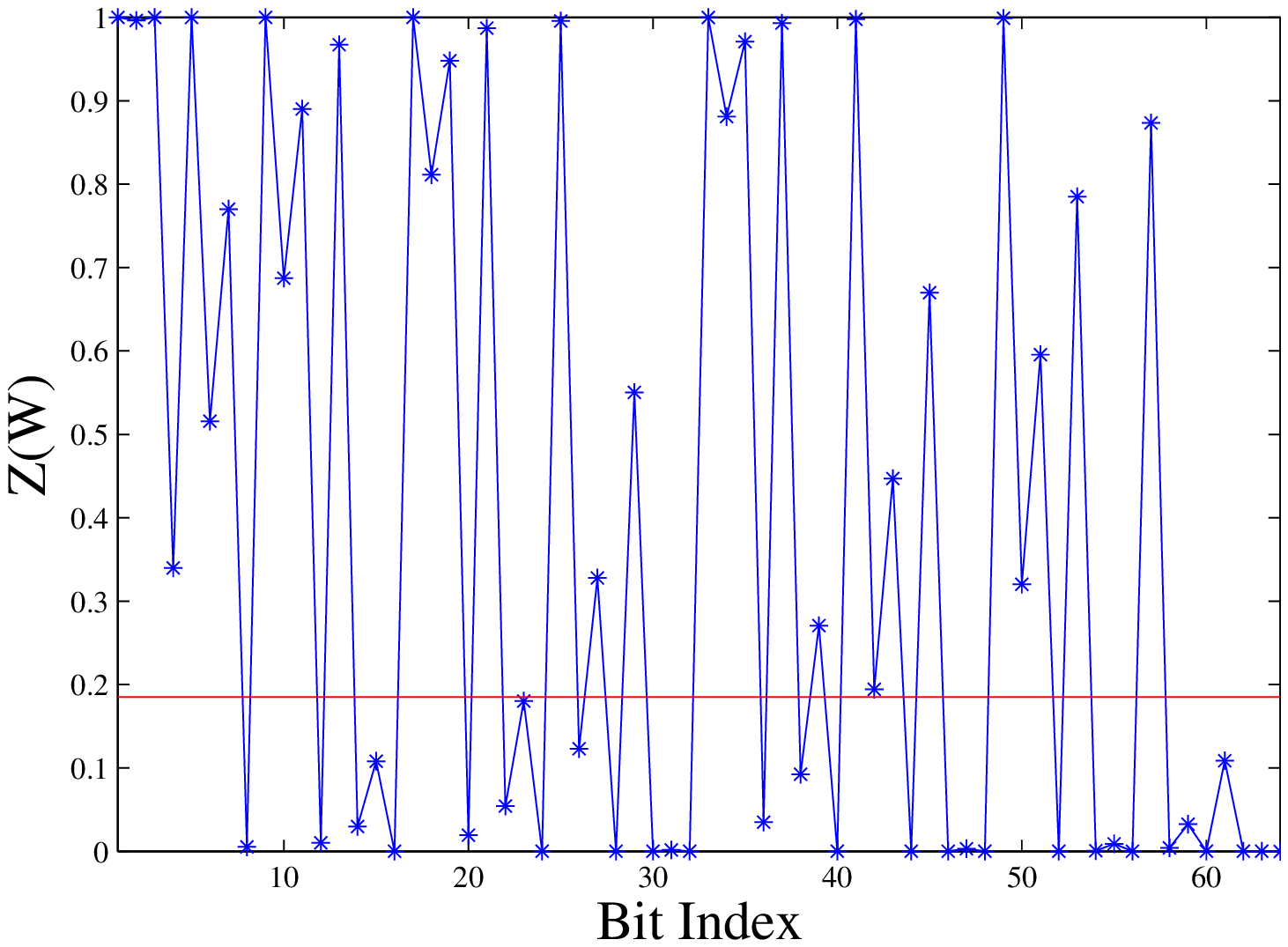}
\par\end{centering}
}\subfloat[Erasure rate $=0.5$]{\begin{centering}
\includegraphics[width=0.65\columnwidth]{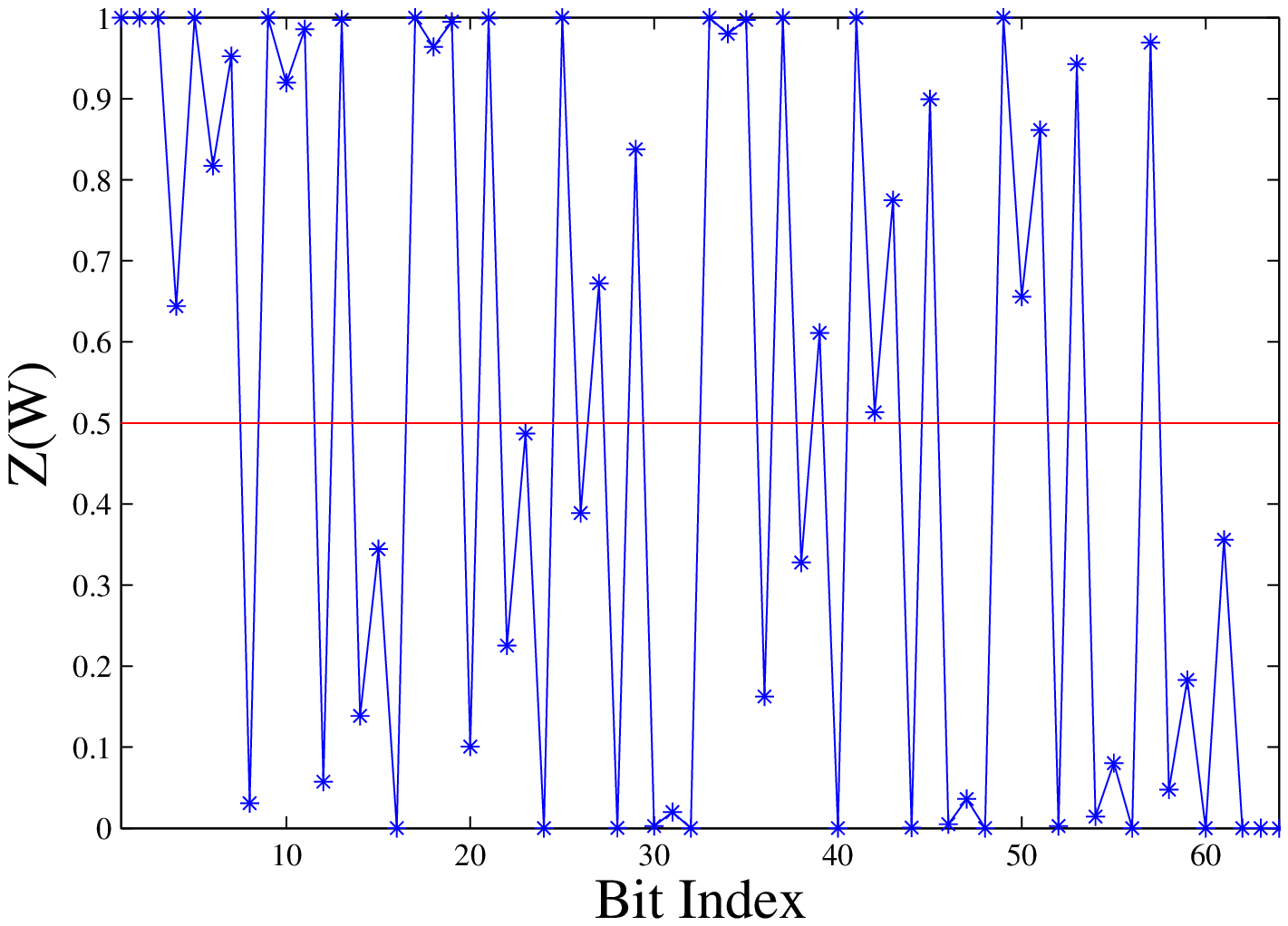}
\par\end{centering}
}\subfloat[Erasure rate $=0.6$]{\begin{centering}
\includegraphics[width=0.65\columnwidth]{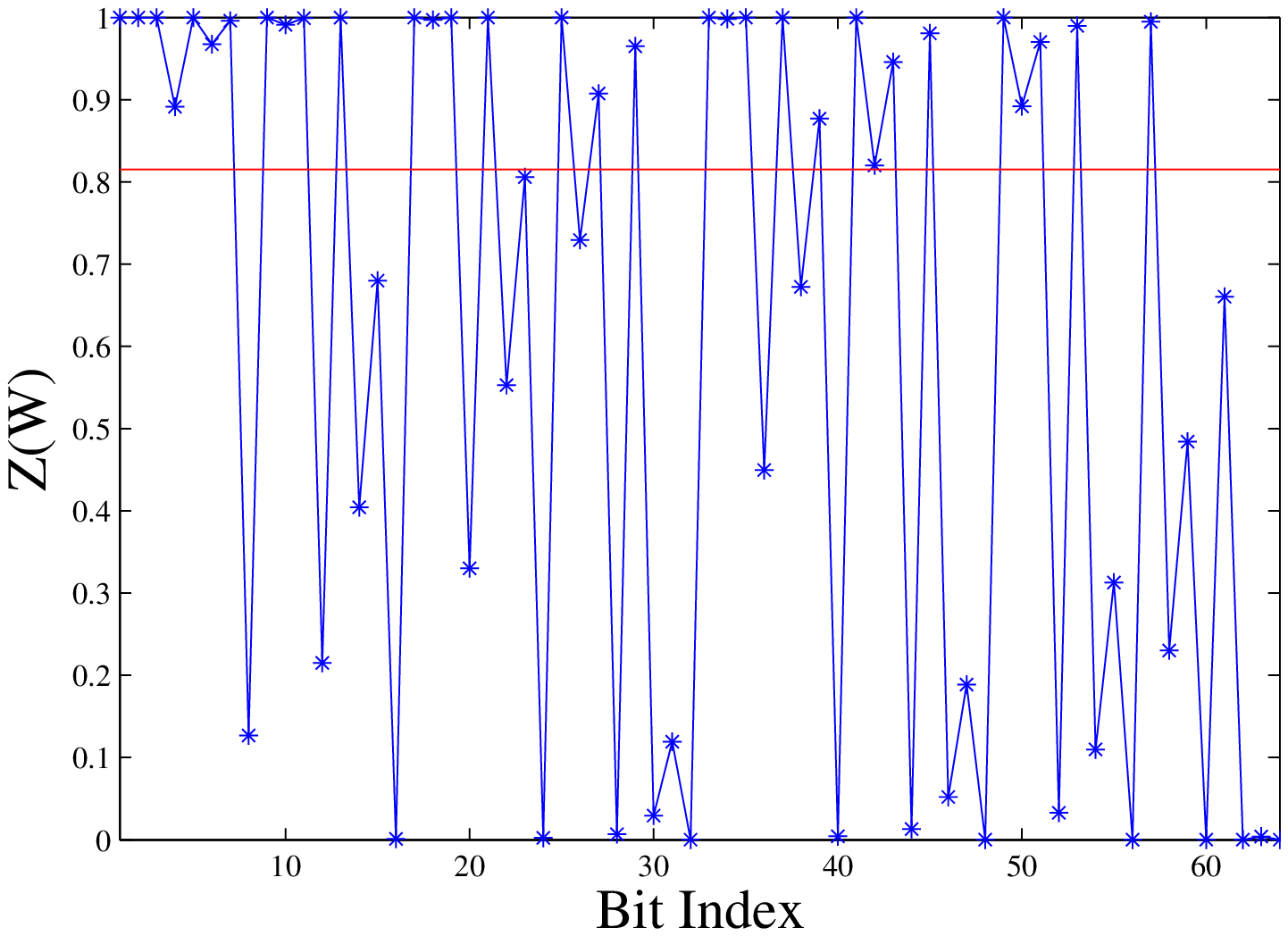}
\par\end{centering}
}

\caption{\label{fig:Bhattacharyya-parameter-versus}Bhattacharyya parameter
versus the bit index for the conventional heuristic method proposed
by Arikan when $N=64$, and half code-rate. The information channels
are below the red lines while the frozen channels are above the red
lines.}
\end{figure*}
where $\sigma^{2}$ is the AWGN channel variance. Therefore, by applying
(\ref{eq:BhatCont}) in the Middleton class-A channel, the Bhattacharyya
parameter can be given in (\ref{eq:BhataIN}). However, there is no
direct solution to (\ref{eq:BhataIN}); hence, the numerical integration
is a feasible solution for finding the $Z(W)$ values in the Middleton
class-A channels.

\section{The Construction of polar codes}

\subsection{Heuristic method}

Let $Z(W_{N}^{i})$ denotes the Bhattacharyya parameter of the bit-channel
$W_{N}^{i}$. The recursive channel parameters for any B-DMC channel
could be calculated as given in \cite{PolarCode}

\begin{eqnarray}
Z(W_{N}^{2i-1}) & \leq & 2Z(W_{N/2}^{i})-(Z(W_{N/2}^{i}))^{2},\label{eq:BhataFirstPart}\\
Z(W_{N}^{2i}) & = & (Z(W_{N/2}^{i}))^{2},\label{eq:BhataSecondPart}
\end{eqnarray}
where $i$ is an integer number $1\leq i\leq N$. In BEC, (\ref{eq:BhataFirstPart})
has equality status which leads to the fact that this construction
could be applied under the BEC only. Consequently, for other channels
such as AWGN, the author of \cite{ArikanletterRMandPC} proposed the
heuristic method in which any channel is treated as an equivalent
BEC with an initial $Z(W_{1})=0.5$. The criteria of this method is
to select the channels with smaller $Z(W)$ for sending information
bits, while select the channels with higher $Z(W)$ as frozen bits.
Due to its simplicity, this method was adopted by a lot of researchers
on polar codes.

To highlight this idea, suppose that $N=64$ and the code rate is
$0.5$. Fig. \ref{fig:Bhattacharyya-parameter-versus} depicts $Z(W)$
of each single channel bits versus the bit index for three cases of
erasure rate $\epsilon=0.4$, $0.5$ and $0.6$. Hence, $32$ channel
bits with the higher $Z(W)$ values are selected as frozen bits, which
are above the red line. In contrast, $32$ channel bits with the lower
$Z(W)$ are selected as information bits, which are below the red
line. It can be noticed that there is a symmetry between the right
side and left side of the figures, especially when $\epsilon=0.5$
. This can be attributed to the fact that for each channel bit $i:i\in N$
with $Z(W_{N}^{i})$, there is another channel bit $j:j\in N$ with
a Bhattacharyya parameter $Z(W_{N}^{j})=1-Z(W_{N}^{i})$. Furthermore,
it can be seen that when the erasure rate changes, the values of $Z(W)$
are changed. Thus, for $\epsilon=0.4$, the red line has a lower value
which means that the $32$ information bits has more reliabilities
due to their Bhattacharyya parameters. In contrast, when $\epsilon=0.6$,
it can be noticed that the information bits have higher values of
Bhattacharyya parameters, i.e., less reliabilities. 

Hence, the heuristic method depends on the properties of the BEC channel
only and it is not an optimum method for other channels such as AWGN
and Middleton class-A. However, this method can give accepted results
when it is used with other channel types. What is interesting in this
method is that its simplicity compared to other methods. In addition,
it is independent of the channel SNR values. Hence, the code designer
needs to apply it according to the codeword length and number of information
bits regardless the channel parameters.

\subsection{Heuristic based capacity method}

\begin{figure*}[tbh]
\subfloat[SNR$=$$2$ dB ]{\begin{centering}
\includegraphics[width=0.65\columnwidth]{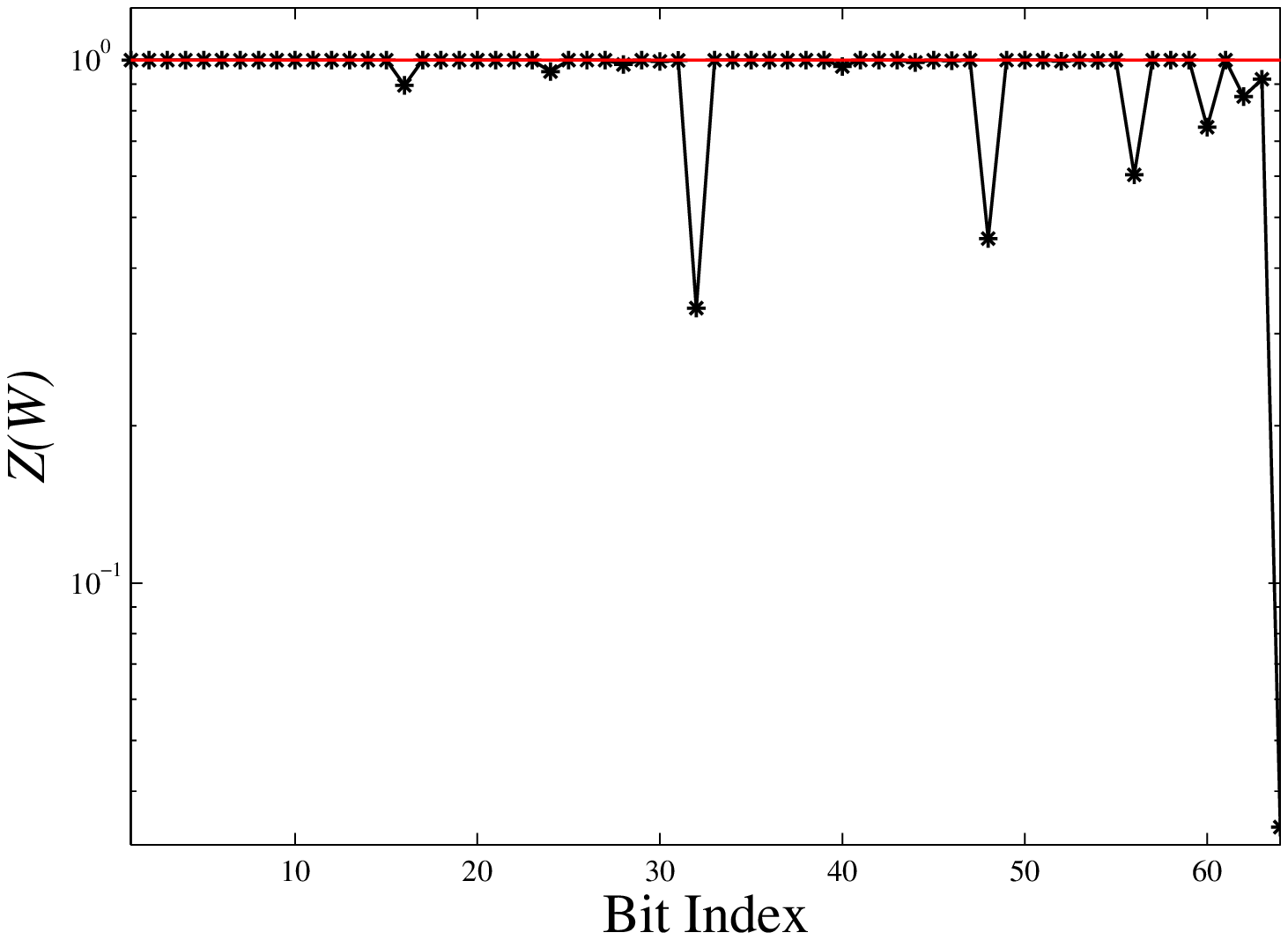}
\par\end{centering}
}\subfloat[SNR$=10$ dB]{\begin{centering}
\includegraphics[width=0.65\columnwidth]{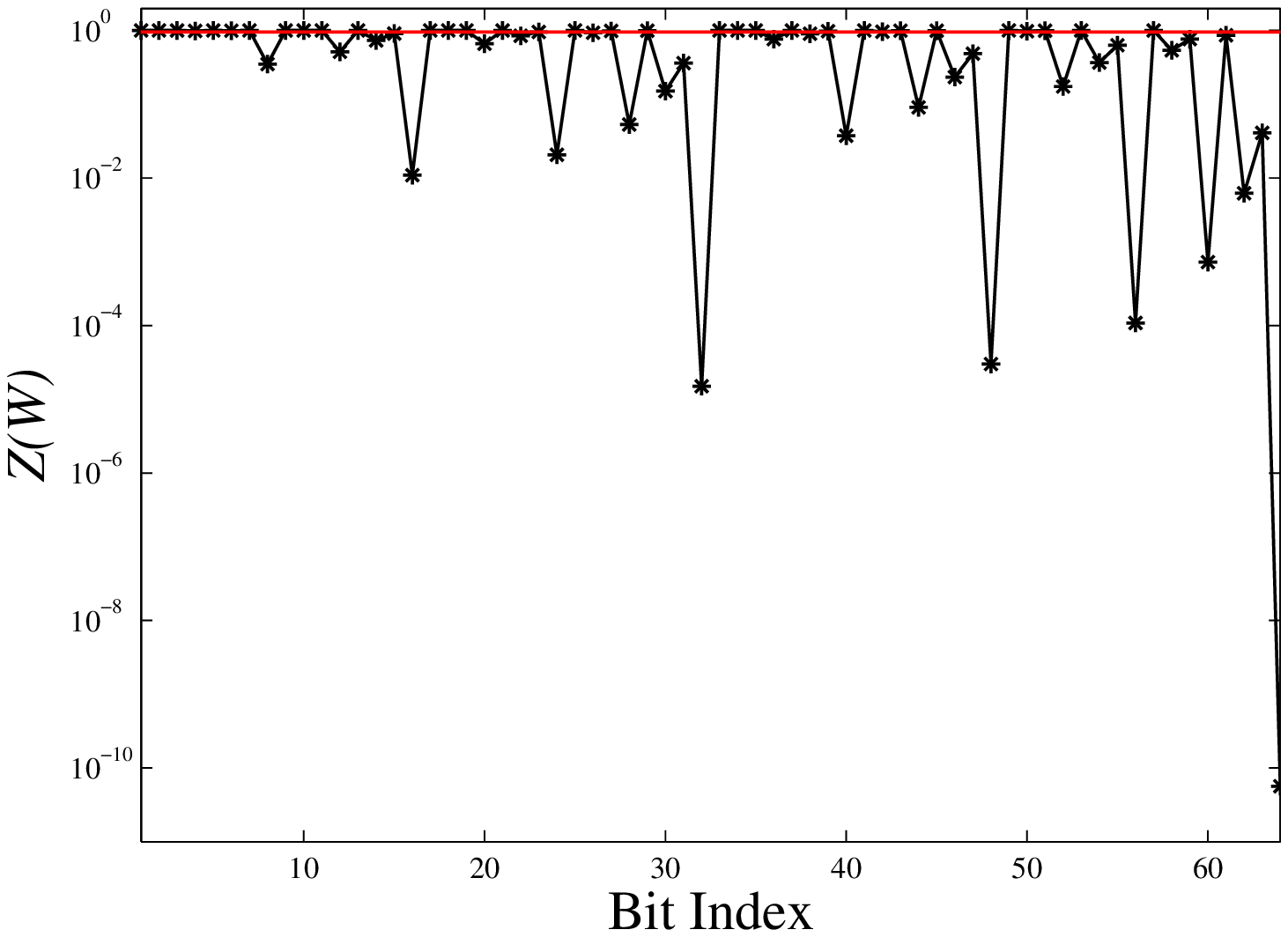}
\par\end{centering}
}\subfloat[SNR $=15$ dB]{\begin{centering}
\includegraphics[width=0.65\columnwidth]{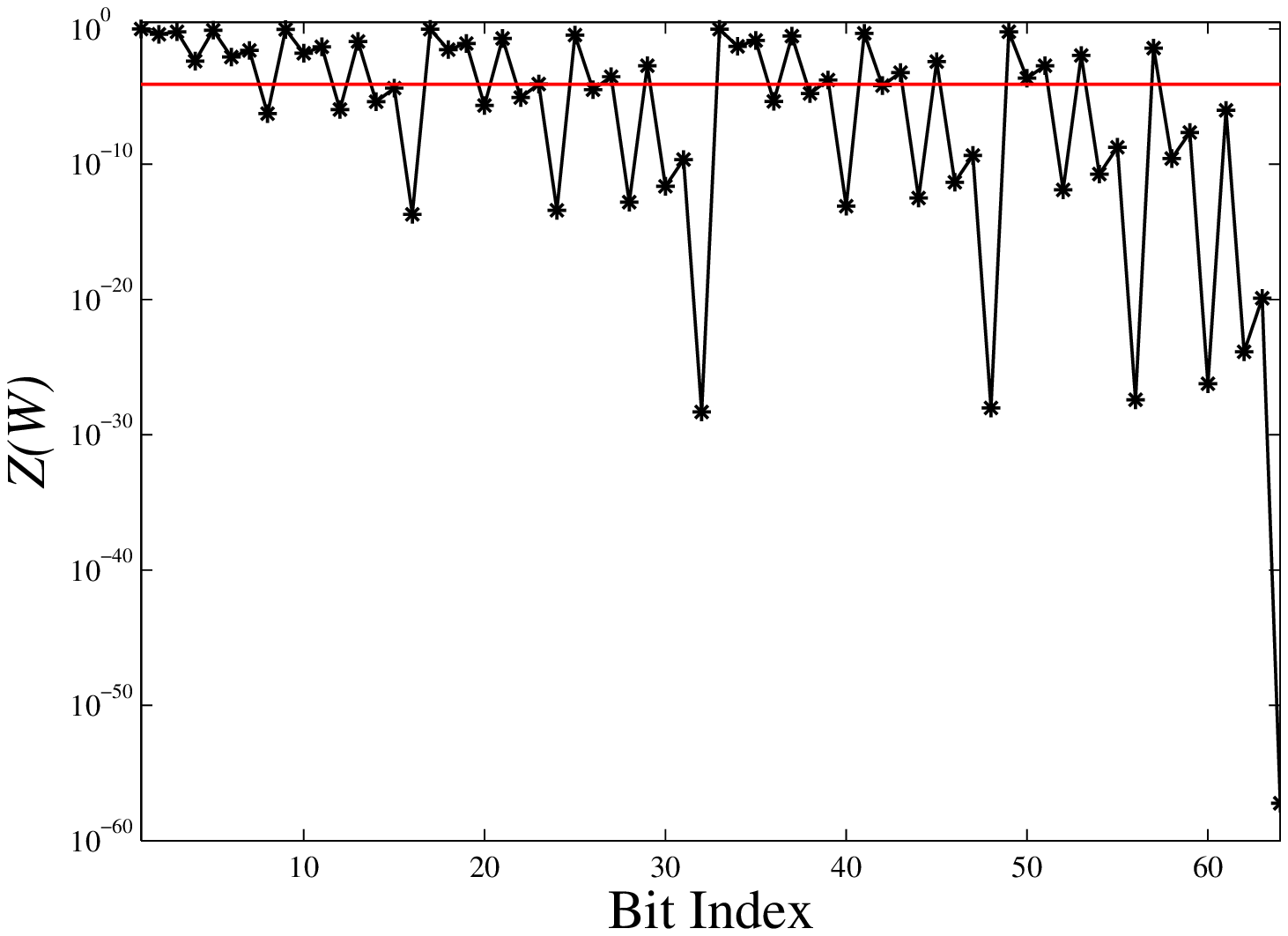}
\par\end{centering}
}

\caption{\label{fig:Bhattacharyya-parameterVSCapacityHeuristic}Bhattacharyya
parameter versus the bit index for the heuristic based capacity method
Arikan when $N=64$, and half code-rate for three different values
of SNR:(a) $2$ dB, (b) $10$ dB, and (c) $15$ dB.}
\end{figure*}

\begin{figure*}[tbh]
\subfloat[BER for channel bits when $E_{b}/N_{0}=2$ dB]{\begin{centering}
\includegraphics[width=0.65\columnwidth]{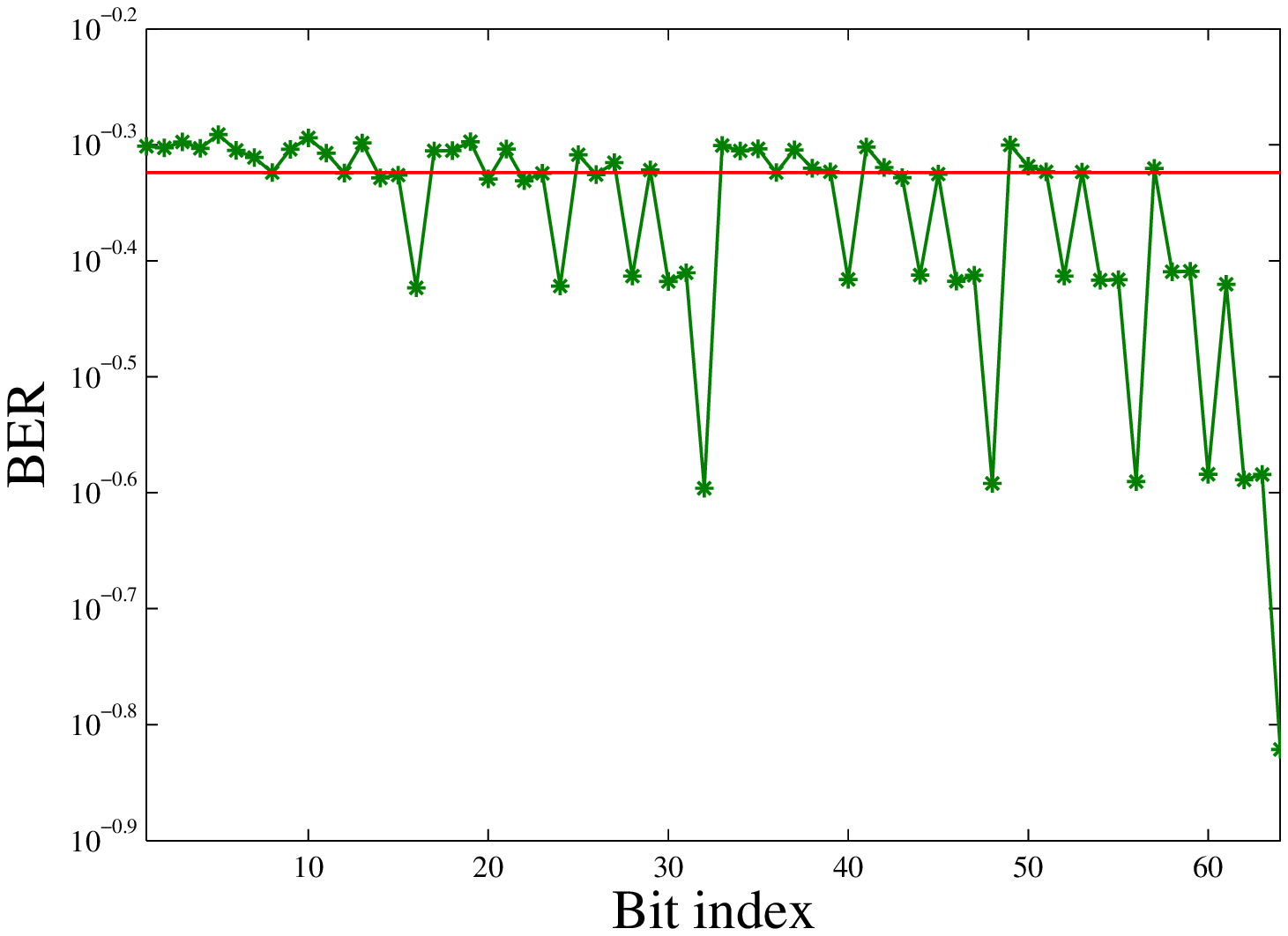}
\par\end{centering}
}\subfloat[BER for channel bits when $E_{b}/N_{0}=10$ dB]{\begin{centering}
\includegraphics[width=0.65\columnwidth]{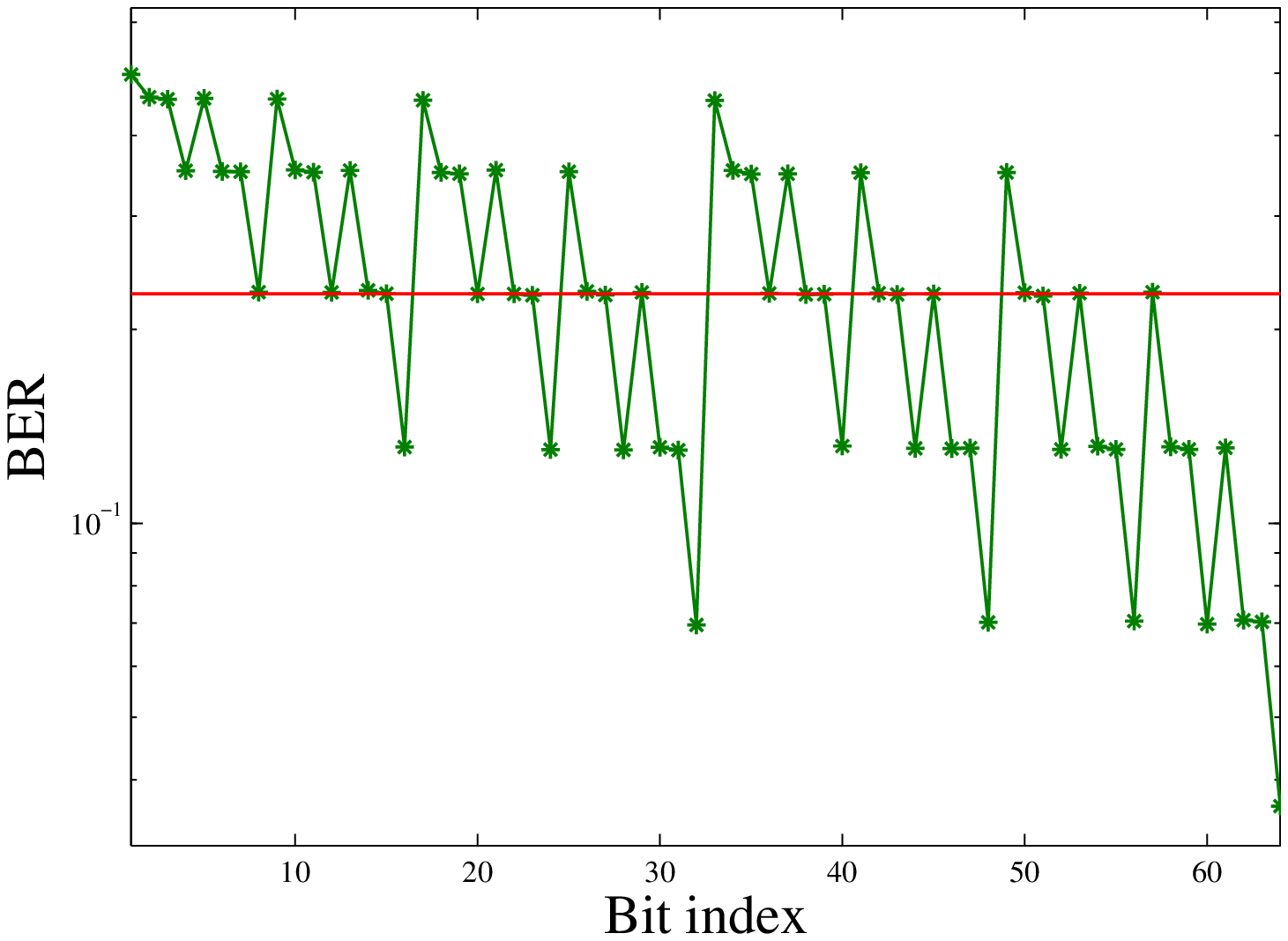}
\par\end{centering}
}\subfloat[BER for channel bits when $E_{b}/N_{0}=15$ dB]{\begin{centering}
\includegraphics[width=0.65\columnwidth]{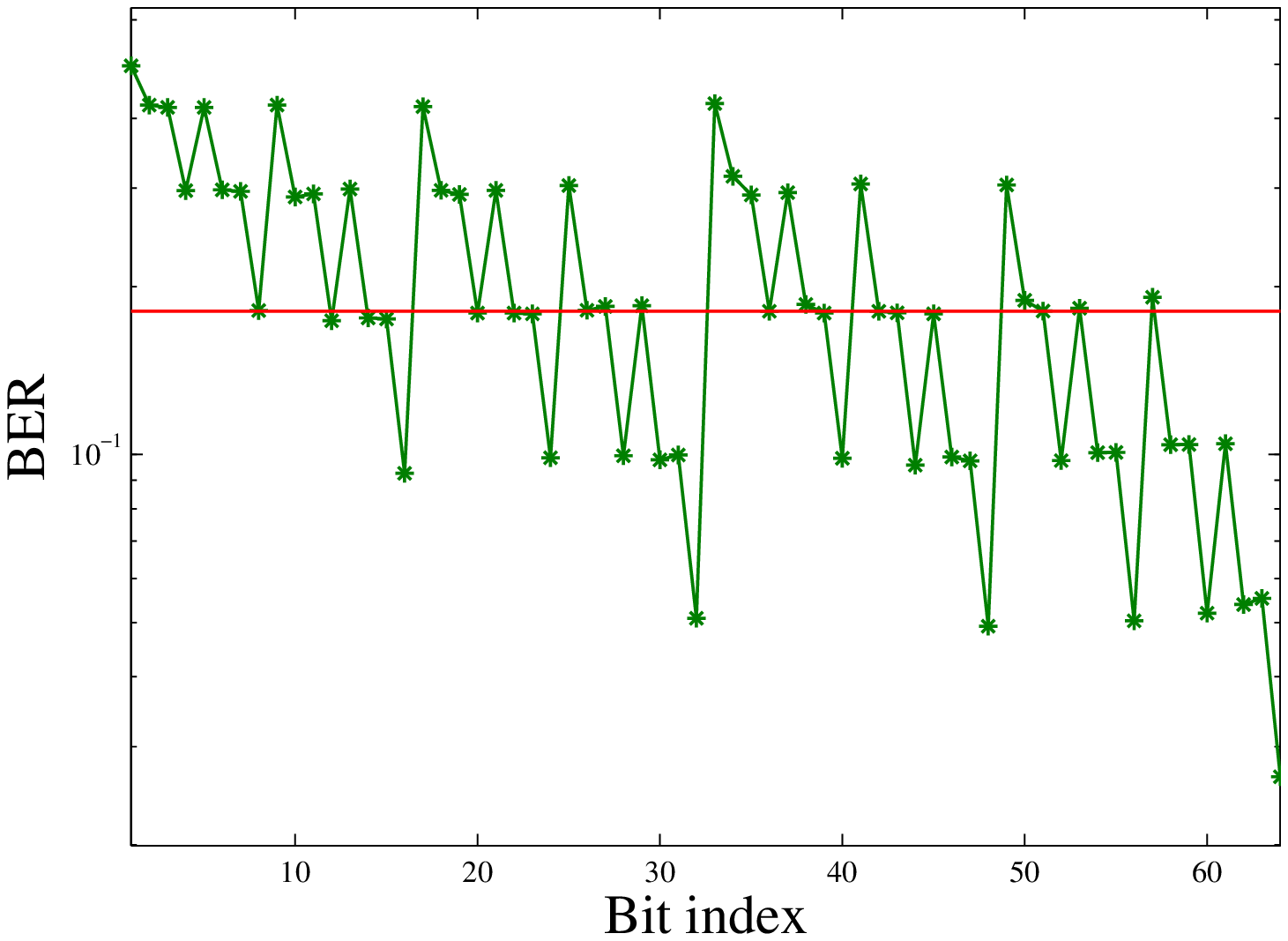}
\par\end{centering}
}

\caption{\label{fig:The-BER-forMC}The BER for channel bits by using the Monte
Carlo method on polar codes where the Middleton class-A channel is
adopted when $N=64$ for three different values of SNR:(a) $2$ dB,
(b) $10$ dB, and (c) $15$ dB..}
\end{figure*}

The algorithm of this method uses (\ref{eq:BhataFirstPart}) and (\ref{eq:BhataSecondPart})
which are used in the normal heuristic method but the major difference
is the initial value which is considered always as $0.5$ in the heuristic
method; whereas in this method, the initial value is evaluated based
on the Bhattacharyya parameters of the channel. Hence, this method
could be regarded as a channel selective method. For example, the
initial value for AWGN is applied according to (\ref{eq:Z(W)forAWGN}).

In this framework, we apply the heuristic based capacity method for
finding the construction of polar codes under Middleton class-A channel
by using (\ref{eq:BhataIN}). Hence, the SNR change affects the construction
in this method. On other words, the optimum code construction for
a specific value of SNR may not be an optimum construction for other
values. Therefore, this method needs to apply on each specific value
of SNR. 

For more clarity, Fig. \ref{fig:Bhattacharyya-parameterVSCapacityHeuristic}
depicts the Bhattacharyya parameters for channels bits versus the
channel index for some values of SNR values. As expected, the reliabilities
of the channel bits are increased with the increasing of the SNR values
as it can be seen that the $Z(W)$ values of the channel bits are
reduced with SNR increasing.

\subsection{Monte Carlo method}

The metric of choosing the best channels is related to the bit error
rate (BER) of the channels as \cite{PolarCode}

\begin{equation}
BER_{i}=Z(W_{N}^{i}).
\end{equation}

Hence, the BER for the bit channels can be evaluated by simulations
with the help of the statistical techniques. Since the fact that there
is no direct algorithm for finding the optimum construction of polar
codes under Middleton class-A parameters, we can adopt the Monte Carlo
simulation for the channel-bits selection. The Monte Carlo is an approximation
statistical method uses samples from $U$, $Y$ as inputs and the
output is the BER values. The values of BER here play the same role
of the $Z(W)$ values in the previous methods. Hence, we can consider
the bits with the highest BER values as frozen bits. 

Fig. \ref{fig:The-BER-forMC} shows the BER for the channel bits when
$N=64$. It is clearly seen that each SNR gives different values of
BER which resulted in different construction due to the changes in
the information set. It can be also noticed that the bits have different
reliabilities; hence, the construction of polar codes selects the
bits with the lowest values of BERs to become the information bits.
The size of information set depends on the code design and in general
when the size increases, it is expected that less reliabilities bits
are included in the information set. The disadvantage of this method
is the huge complexity compared to other methods.

\section{Simulation Results and Discussions }

\begin{figure*}[tbh]
\subfloat[SNR $=2$ dB]{\includegraphics[width=0.65\columnwidth]{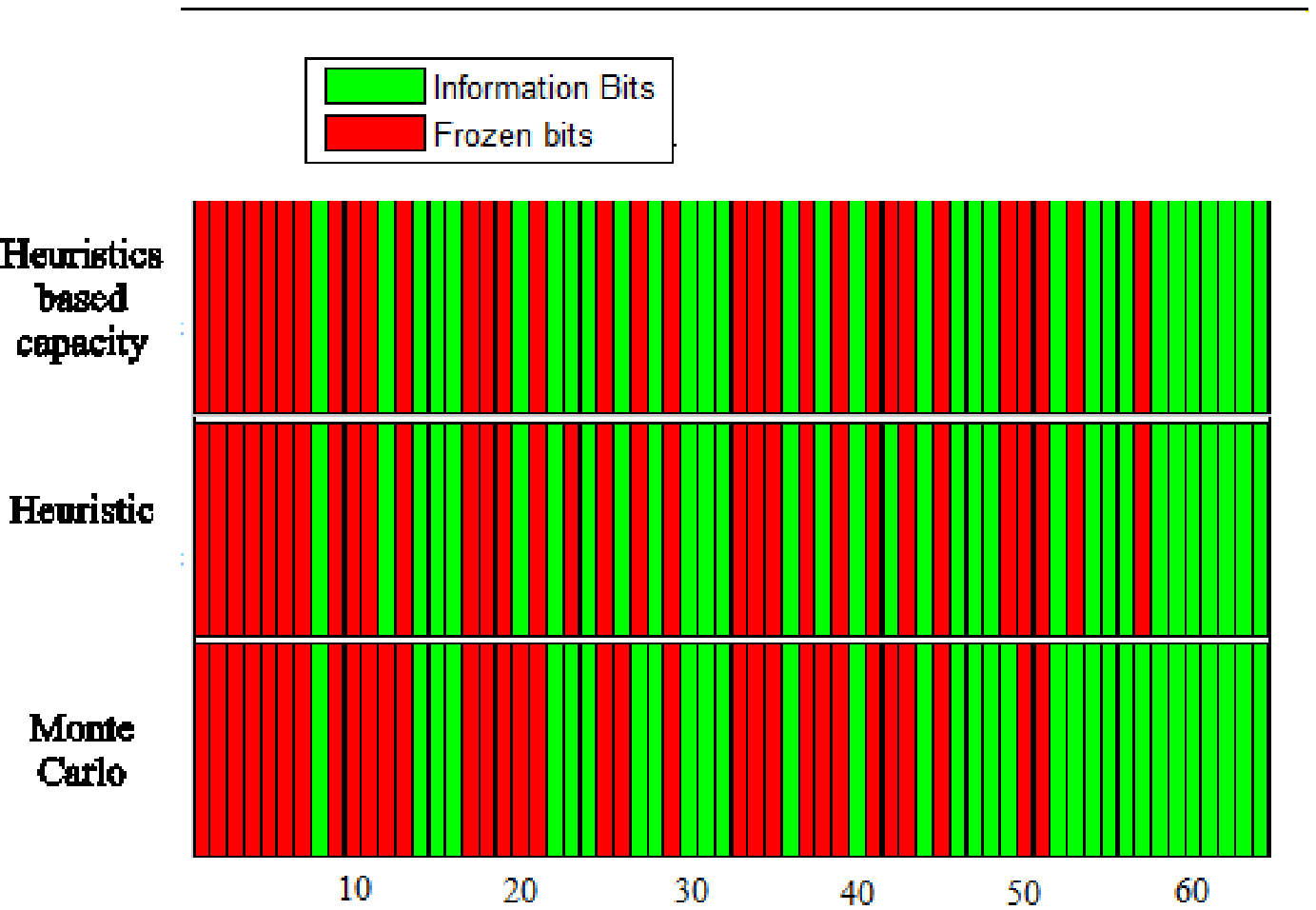}

}\subfloat[SNR $=5$ dB]{\includegraphics[width=0.65\columnwidth]{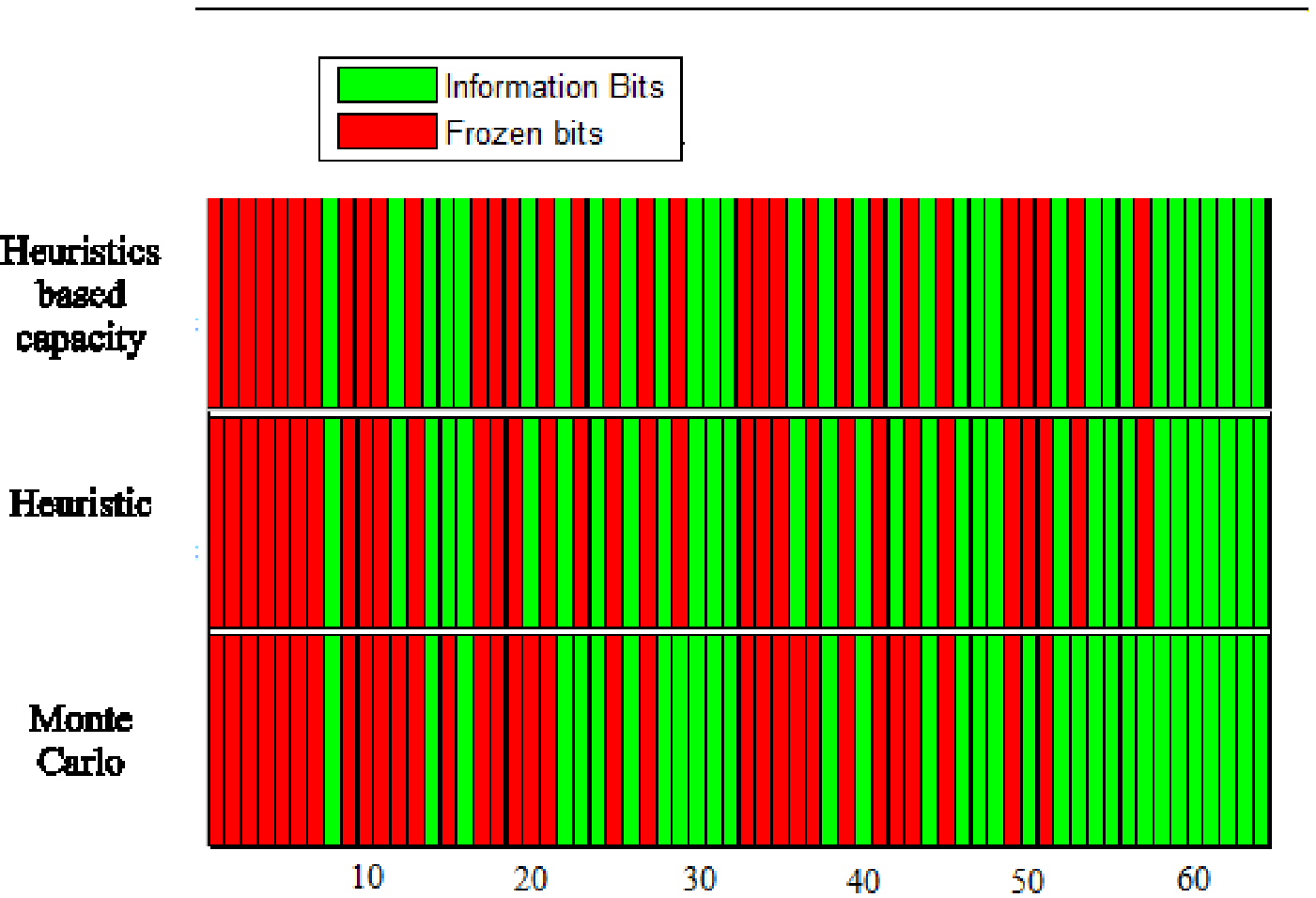}

}\subfloat[SNR $=10$ dB]{\includegraphics[width=0.65\columnwidth]{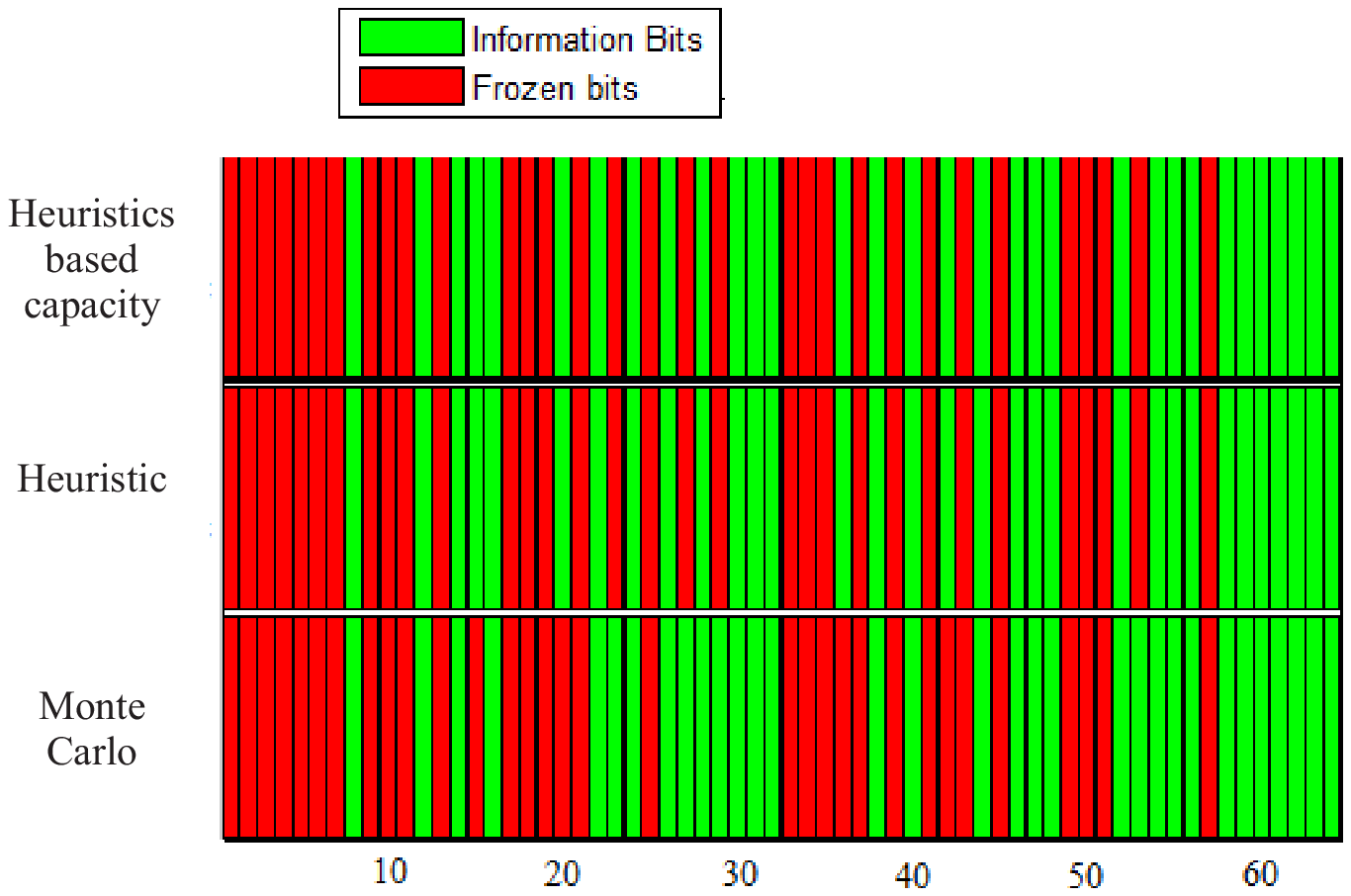}

}

\caption{\label{fig:ThePCConstructionofthe3Methods}The construction of polar
codes with the three methods where the first row is heuristic based
capacity, second row is heuristic method and third row is the Monte
Carlo method when $N=64$, and half code-rate for three different
values of SNR:(a) $2$ dB, (b) $5$ dB, and (c) $10$ dB}
\end{figure*}

In this section, we investigate the construction of the three methods
in different values of SNR. Fig. \ref{fig:ThePCConstructionofthe3Methods}
illustrates the three methods of construction where the code length
is $64$ and the code-rate is $0.5$ for three values of SNR. The
construction in this case is responsible for selecting $32$ bits
from the entire codeword for carrying the information and $32$ bits
as frozen bits. The first row in the figure represents the heuristic
based capacity method, the second row represents the conventional
heuristic method and finally the third row represents the Monte Carlo
method. It can be seen that the heuristic method is not affected by
the change of the SNR values, while the two other methods can be affected
by SNR. Although the constructions look similar for the three methods,
the slight differences in SNR may affect the performance. It is worth
mentioning that these differences are increased with the increasing
of the code length. 

\begin{figure}[t]
\begin{centering}
\includegraphics[width=1\columnwidth]{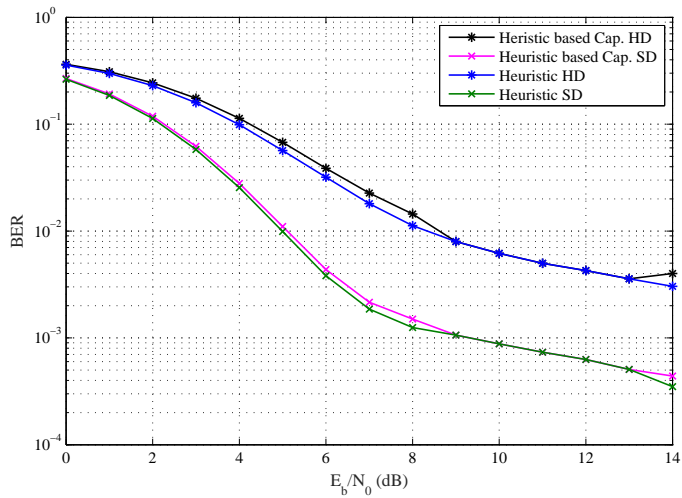}
\par\end{centering}
\caption{\label{fig:FER-performance-versusN64}BER performance versus $E_{b}/N_{0}$
for both heuristic and heuristic based capacity methods when $N=64$
and half code-rate.}
\end{figure}

\begin{figure}[tbh]
\begin{centering}
\includegraphics[width=1\columnwidth]{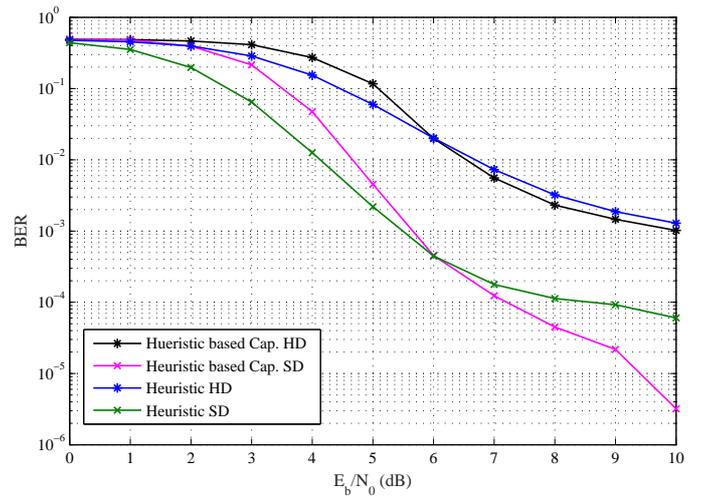}
\par\end{centering}
\caption{\label{fig:FER-performance-N512}BER performance versus $E_{b}/N_{0}$
for both heuristic and heuristic based capacity methods when $N=512$
and half code-rate.}
\end{figure}

For better clarity, we compared the performance of the code under
only heuristic and heuristic based capacity methods due to the huge
complexity which is needed for the Monte Carlo method. Hence, a comparison
between these two methods in terms of BER performance is illustrated
in Fig. \ref{fig:FER-performance-versusN64} when $N=64$ and Fig.
\ref{fig:FER-performance-N512} when $N=512$. It can be noticed that
with a small size of codeword, the heuristic based capacity method
performance is similar to the heuristic method performance. In contrast,
when $N=512$, the heuristic based capacity method outperforms the
normal heuristic method with larger $E_{b}/N_{0}$. This can be attributed
to the positive impact of the polarization phenomenon on the heuristic
based capacity method as it is known that the polarization is increased
with the increasing in the code size. Hence, the construction with
the heuristic based capacity method becomes more useful with large
code sizes while for the small and moderate codewords, both methods
give similar performance. 

\section{Conclusion}

We investigated in this paper the construction of polar codes under
the impact of the impulsive noise in PLC systems. We focused on three
methods of code construction which are heuristic, heuristic based
capacity and Monte Carlo methods. Furthermore, we showed a comparison
of the performance between the heuristic and heuristic based capacity
methods in terms of the BER behavior. Note that the successive cancellation
decoder is adopted in both hard-decision (HD) and soft-decision (SD)
in the comparison. We found that there are slight differences between
these methods with moderate and small block length but when the codeword
is increased, the heuristic based capacity method outperforms the
performance of the conventional heuristic method. 

\bibliographystyle{IEEEtran}
\bibliography{2yearReoperrefernces}

\end{document}